\documentclass[journal]{IEEEtran}
\usepackage{amssymb}
\usepackage{makecell}
\usepackage{multirow}
\usepackage{colortbl}
\ifCLASSINFOpdf
\else
\fi

\usepackage{float}
\usepackage{graphicx}
\usepackage{hyperref}
\usepackage{cite}
\usepackage{enumerate}
\usepackage{pifont}
\usepackage{color}
\usepackage{bbding}
\usepackage{amsmath}

\usepackage{amssymb}
\usepackage{mathrsfs,balance}
\usepackage{mathtools}
\usepackage{pifont}
\usepackage{bbding}
\usepackage{amsmath}
\usepackage{amssymb}
\usepackage{mathrsfs}
\usepackage{graphicx}
\usepackage{amsthm}
\usepackage{subfigure,balance}
\usepackage{amssymb, amsmath, amsfonts, epsfig, latexsym, times}
\usepackage{bm}
\usepackage{cite}
\usepackage{algorithm}
\usepackage{algorithmic}
\usepackage{setspace}
\usepackage{url}
\usepackage{booktabs}
\usepackage{amssymb}
\usepackage{verbatim}
\usepackage{stfloats}
\IEEEoverridecommandlockouts
\def\BibTeX{{\rm B\kern-.05em{\sc i\kern-.025em b}\kern-.08em
    T\kern-.1667em\lower.7ex\hbox{E}\kern-.125emX}}

\begin{document}
	\title{{Sum Secrecy Rate Maximization for IRS-aided Multi-Cluster MIMO-NOMA Terahertz Systems}}
	\author{\IEEEauthorblockN{Jinlei Xu, Zhengyu Zhu, \emph{Senior Member, IEEE}, Zheng Chu, \emph{Member, IEEE}, Hehao Niu, \emph{Member, IEEE}, \\Pei Xiao, \emph{Senior Member, IEEE}, and Inkyu Lee,~\emph{Fellow, IEEE}		
		}
		\thanks{
		This work was supported in part by the open research fund of National Mobile Communications Research Laboratory Southeast University under Grant 2023D11; in part by Sponsored by Program for Science \& Technology
		Innovation Talents in Universities of Henan Province under Grant 23HASTIT019; in part by Natural Science Foundation of Henan Province under Grant 232300421097; in part by the Project funded by China Postdoctoral Science Foundation under Grant 2020M682345; in part by the Henan Postdoctoral Foundation No.202001015; in part by the National Research Foundation of Korea funded by the Ministry of Science and ICT, Korea Government under Grant 2017R1A2B3012316. \emph{(Corresponding author: Zhengyu Zhu)} }
		\thanks{J. Xu is with the School of Electrical and Information Engineering, Zhengzhou University,
			Zhengzhou 450001, China. (e-mail: iejlxu@mail.dlut.edu.cn).
			
			Z. Zhu is with the School of Electrical and Information Engineering, Zhengzhou University, Zhengzhou, 450001, China, and is~also with National Mobile Communications Research Laboratory, Southeast University, Nangning 210018, China (e-mail: zhuzhengyu6@gmail.com)
			
			Z. Chu and P. Xiao are with the 5GIC \& 6GIC, Institute for Communication Systems (ICS), University of Surrey,
			Guildford GU2 7XH, UK. (e-mail: andrew.chuzheng7@gmail.com, p.xiao@surrey.ac.uk)
			
			H. Niu is with the Institute of Electronic Countermeasure, National University of Defense Technology, Hefei 230037, China. (e-mail:
			niuhaonupt@foxmail.com)		
			
			I. Lee is with School of Electrical Engineering, Korea University, Seoul, Korea. (e-mail: inkyu@korea.ac.kr)
		}
		\vspace{-30pt}
	}	
	\maketitle

\begin{abstract}
Intelligent reflecting surface (IRS) is a promising technique to extend the network coverage and improve spectral efficiency. This paper investigates an IRS-assisted terahertz (THz) multiple-input multiple-output (MIMO)-nonorthogonal multiple access (NOMA) system based on hybrid precoding with the presence of eavesdropper. Two types of sparse RF chain antenna structures are adopted, i.e., sub-connected structure and fully connected structure.  First, cluster heads are selected for each beam, and analog precoding based on discrete phase is designed. Then, users are clustered based on channel correlation, and NOMA technology is employed to serve the users. In addition, a low-complexity forced-zero method is utilized to design digital precoding in order to eliminate inter-cluster interference.
On this basis, we propose a secure transmission scheme to maximize the sum secrecy rate by jointly optimizing the power allocation and phase shifts of IRS subject to the total transmit power budget, minimal achievable rate requirement of each user, and IRS reflection coefficients. Due to multiple coupled variables, the formulated problem leads to a non-convex issue. We apply the Taylor series expansion and semidefinite programming to convert the original non-convex problem into a convex one. Then, an alternating optimization algorithm is developed to obtain a feasible solution of the original problem. Simulation results verify the convergence of the proposed algorithm, and deploying IRS can bring significant beamforming gains to suppress the eavesdropping.

\end{abstract}
\begin{IEEEkeywords}
Intelligent reflecting surface, terahertz, MIMO-NOMA, hybrid precoding, secure communication.
\end{IEEEkeywords}
\IEEEpeerreviewmaketitle

\section{Introduction}

With the accelerated progress of the global round of new technological revolution and industrial transformation, the widespread adoption and application of data-centered technologies such as artificial intelligence (AI), virtual reality, autonomous driving, and smart cities have posed significant challenges to the fifth-generation mobile communication (5G) networks\cite{Alwis, Wang6}. Therefore, it is imperative to develop the sixth-generation mobile communication (6G) technologies with new characteristics for the communication demands of the future information society \cite{Wang1, Chowdhury1}.

To meet the development vision of "better performance, greater intelligence, broader coverage, and greener," 6G puts forward higher requirements for a series of indicators such as data rates, transmission latency, and secure communication \cite{You1}. Compared with 5G, 6G can continuously expand the breadth and depth of applications in the mobile Internet of Things and enable seamless connectivity among all things, which will better supports the robust development of the digital economy \cite{Karam1}.
Conventional wireless communication systems operating below 5 GHz, and even the millimeter-wave (mmWave) technology are unable to meet the massive connectivity demands of 6G. Therefore, new wireless technologies are urgently needed to address the challenge of capacity shortage\cite{Guangyi1}.
Compared with the current wireless technologies, terahertz (THz) communication possesses higher frequencies and wider bandwidth, enabling support for a greater number of connected devices and achieve high transmission rates from 100~Gbit/s to 1~Tbit/s \cite{Hang1}. It can alleviate the issues of spectrum scarcity and capacity limitations in current wireless systems and is considered one of the potential technologies for 6G.

Due to the inherent broadcast nature of wireless channels, wireless systems are vulnerable to malicious attack, phishing and eavesdropping \cite{Zhaonan1}.
In order to deal with the serious threat to the secure transmission of the communication system, the research on the physical layer security (PLS) of wireless communications by leveraging the channel propagation characteristics has gradually become a research hotspot in the field of wireless security \cite{Zhengyu1}.

Recently, intelligent reflecting surface (IRS), composed by a large number of low-cost meta-material units, can dynamically configure the propagation environment of wireless signals via programming to improve the comprehensive performance of systems, which has been considered as one of the promising technologies of 6G \cite{Zheng2}. 
In particular, by adaptively adjusting the reflection coefficients and amplitudes of IRS elements, a directional energy beam can be formed and reflected to the legitimate receivers, which can effectively suppress eavesdropping and expand the signal coverage\cite{Zhengyu2,Qingqing1}.
IRS can be used to mitigate path loss and blockage issues in THz system. Deploying IRS can bypass obstacles to provide a virtual line-of-sight (LoS) link for THz communications, thus alleviating signal attenuation caused by molecular absorption and improving the security of transmission \cite{Chenzhi1}. An IRS assisted THz indoor communication system is studied in \cite{Yijin1}, by jointly optimizing IRS deployment and reflection coefficients, THz sub-band allocation, as well as power control to maximize sum rate of users. In \cite{Jinlei1}, an IRS assisted simultaneous wireless information and power transmission THz security system was proposed when the LoS link is blocked. Under the outage probability constraints, the problem of jointly optimizing the transmit beamforming and IRS phase shifts to minimize the total transmit power of the system was studied. In addition, deploying IRS and using active and passive beamforming can effectively suppress eavesdropping to improve the security. In \cite{Jingping1}, Qiao $et~al.$ studied the problem of secure communication in THz system assisted by multiple IRSs. Considering the imperfect eavesdropping channel state information (CSI), the worst-case security rate was derived, and a robust secure strategy for multi eavesdropping system was proposed.

On the other hand, non-orthogonal multiple access (NOMA) can support large-scale access and effectively utilize spectrum resources, thus receiving extensive attention from the academic and industrial communities\cite{Xingwang1, Xingwang2}. Compared to the traditional orthogonal multiple access (OMA) scheme, NOMA can significantly improve transmission rates and system capacity by serving multiple users in a single resource block \cite{Yang1}.
In addition, serial interference cancellation (SIC) technology is leveraged at the NOMA receivers to eliminate co-channel interference between users and realize the reuse of limited spectrum resources\cite{Xingwang3} .

Furthermore, combining multiple-input multiple-output (MIMO) precoding with NOMA can significantly enhance system throughput \cite{Xiaoling1}.  In MIMO systems, equipping each antenna with a dedicated radio frequency (RF) chain results in higher energy consumption and hardware costs. Hybrid precoding (HP) adopted sparse RF links can reduce the number of RF chains required by the system, thereby improving energy efficiency without sacrificing performance \cite{Linglong1}. Moreover, there are two typical structures for sparse RF chains: fully connected (FC) and sub-connected (SC) \cite{Jingbo1}.
The spectral efficiency and energy efficiency of two typical hybrid precoding structures in MIMO systems with limited feedback channels were investigated in \cite{Shaojun1}. Among them, the FC structure has higher complexity and energy consumption, resulting in lower energy efficiency compared to the SC structure. Conversely, the SC architecture is simpler and more energy-efficient, but has lower spectrum efficiency \cite{Wanming1}.

The THz possesses abundant spectral resources and exhibits strong directionality, which can ensure the better correlation of channels. Therefore, by combining THz with MIMO-NOMA technology can significantly enhance spectrum efficiency \cite{Hadi1}.
To address the unreliability of THz transmission caused by the heterogeneous networks, Xu $et~al.$ proposed an intelligent reconfigurable THz MIMO-NOMA framework \cite{Xiaoxia1}. This framework allows for flexible configuration of mixed beams to meet the demands of ultra-high data rates and massive connections.
To alleviate the problem of information security and increased power consumption brought about by large-scale network access, Zhang $et~al.$ investigated the problem of maximizing energy efficiency (EE) in THz NOMA-MIMO systems. Based on the channel correlation characteristics, a user clustering algorithm based on enhanced k-means is proposed. In addition, they utilized a HP scheme based on SC to reduce power consumption to maximize the  achievable EE with the presence of incomplete SIC\cite{Haijun1}.

Although many works have been conducted on the optimization of IRS, THz transmission and MIMO-NOMA, the above technologies are rarely optimized jointly, and clustering of system with large-scale multi-users is not considered. This oversight can lead to collisions among nodes sharing the same channel, resulting in a decrease in user service quality. Clustering multiple users can reduce routing overhead and improve network capacity.
Inspired by their respective advantages, integrating them can further improve the security, spectrum efficiency and coverage. This paper considers an IRS-assisted multi-cluster THz MIMO-NOMA system.
Our goal is to maximize the sum secrecy rate (SSR) by jointly optimizing the power allocation and IRS phase shifts while satisfying the total power budget, minimal rate requirement of users and IRS reflection coefficients. The main contributions of this paper can be summarized as follows.


\begin{itemize}
	\item We propose an IRS-assisted multi-cluster THz MIMO-NOMA system, in which the base station (BS) employs a HP architecture with sparse RF chains, including both FC and SC architectures. Moreover, a clustering algorithm based on correlation channel gains of users is employed to mitigate inter-cluster interference.	
	When an eavesdropper involved, an optimization problem is formulated to jointly optimize the power allocation and IRS phase shifts matrix, aiming to maximize the SSR.
	
	\item For the design of hybrid analog and digital precoding, we first perform clustering based on the channel conditions of the users, and select the cluster heads. Then, analog precoding is designed based on the equivalent channels of each cluster head to enhance the antenna array gain. Finally, zero-forcing (ZF) digital precoding is leveraged to eliminate inter-user interference with the maximum gain in the equivalent channels within each cluster.
	
	It is difficult to tackle the original problem directly due to the non-convexity of the objective function. As such, the joint optimization problem is decomposed into two sub-problems, which can be transformed into the tractable convex form via semi-definite relaxation (SDR) and the first-order Taylor expansion. Then, with the obtained analog and digital precoding, the sub-optimal power allocation and IRS phase shifts can be calculated through an alternating optimization (AO) algorithm.
	\begin{figure}[!htbp]
		\centering
		\includegraphics[height=6.5cm,width=8.5cm]{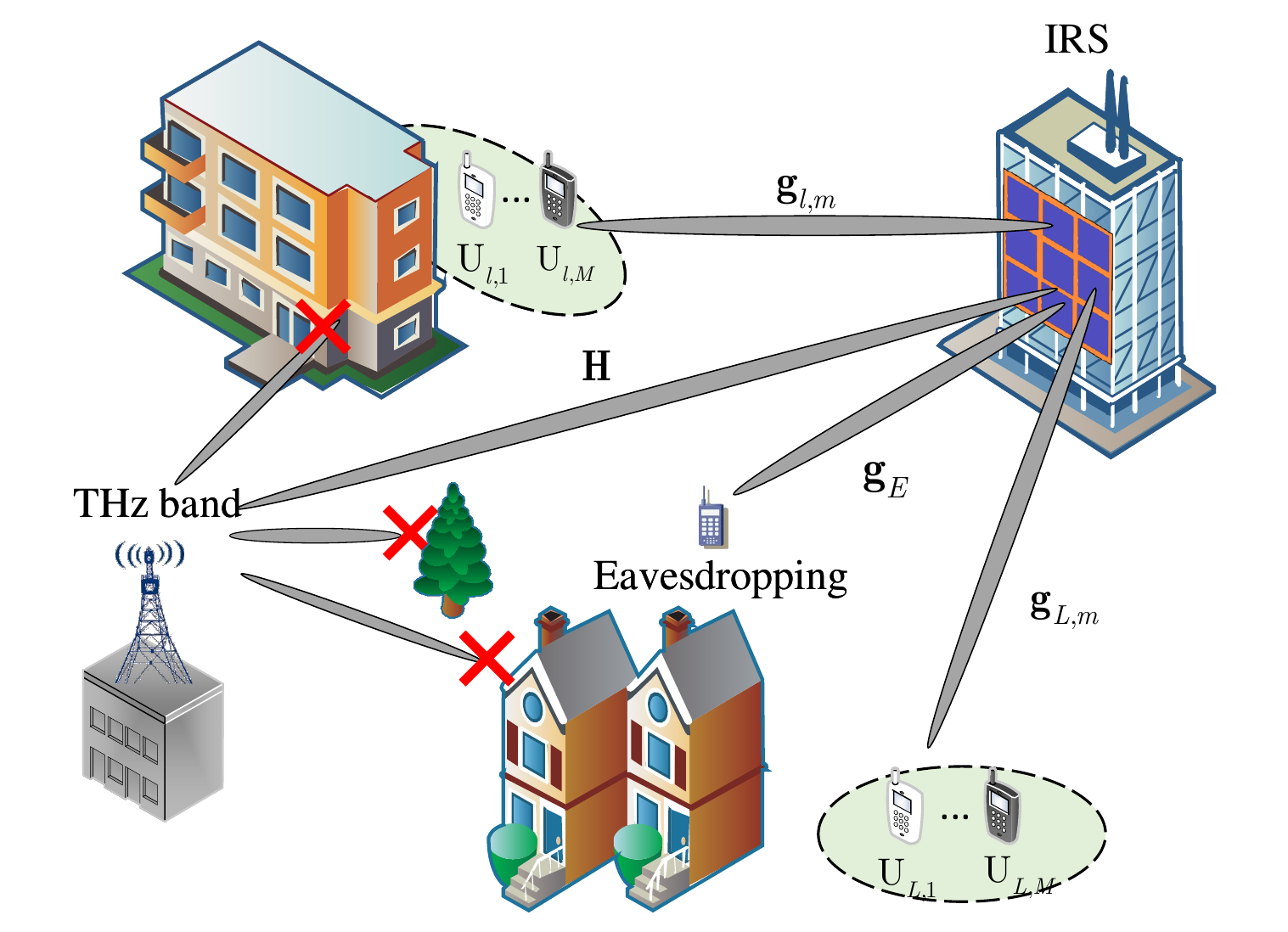}
		\caption{System model.}
		\label{fig:system}
	\end{figure}
	\item The security transmission performance of the proposed IRS-assisted multi-cluster THz MIMO-NOMA system is evaluated through simulations, which shows that the leveraged IRS passive beamforming can effectively suppress the eavesdropping. Additionally, FC architecture achieves higher SSR compared to SC, while having lower secure energy efficiency (SEE). What is more, there is a trade-off between SEE and SSR in the proposed scheme.
	
\end{itemize}

The rest of the paper is organized as follows. Section II introduces the system model of IRS assisted THz MIMO-NOMA network. In Section III, user clustering and hybrid precoding are discussed carefully. In Section IV, we develop an alternate iteration algorithm  to maximize sum secrecy rate under the rate requirement of each user. In Section V, our simulation results demonstrate the effectiveness of the proposed algorithms. Finally, we summarize the paper in Section VI.

\textbf{Notations}: Vectors and matrices are represented by bold lowercase
and uppercase letters, respectively. For a matrix $\bf{A}$, ${(\bf{A})}^T$ and ${(\bf{A})}^H$ represent transpose and Hermitian transpose, respectively. ${\mathbf{A}}\underline  \succ  0$ denotes that ${\mathbf{A}}$ is a semi-positive definite matrix. ${\text{Tr}}\left( {\mathbf{A}} \right)$ stands for the trace of a matrix ${\mathbf{A}}$. 
${\text{diag}}\left( {\mathbf{a}} \right)$ is a diagonal matrix.
${\mathbb{C}^{{\rm{x}} \times y}}$ means the space of $x \times y$ complex matrix.
$\mathrm{E}[\cdot]$ describes expectation. $\left\| \cdot \right\|$ stands for the Euclidean norm of the vector. $\mathcal{CN}(\mu,\sigma^2) $ denotes the a complex Gaussian distribution with mean $\mu$ and variance $\sigma^2$.

\section{System Model and Problem Formulation}
As shown in Fig. 1, we investigate an IRS-aided multi-cluster THz MIMO-NOMA system. In particular, the BS with HP structure is equipped with ${N_{\rm{TX}}}$ antennas, broadcasting the superimposed signal information with the NOMA technique on each beam. 
Since the direct links between the BS and users are blocked by obstacles, i.e., buildings, an IRS, equipped with  ${N_{\rm{IRS}}}$ reflecting elements, passively participates in the secure transmission from the BS to $M$ single antenna legitimate users in the presence of a single-antenna eavesdropper (Eve). Moreover, it is assumed that all users can only receive reflected signals from IRS.

User clustering can help to achieve the good performance for the NOMA system \cite{Shaojun1}, thus we divide users into $L$ groups,
and assume that the number of clusters is the same as the number of RF chains, i.e., $L = {N_{\rm{RF}}}$, and select cluster heads from users according to the channel gain and its correlation. Moreover, the users are grouped according to the scheme in \cite{Linglong1}. By using NOMA, each beam can support multiple users, and the inter-user interference can be eliminated by applying the SIC technique to users with lower channel gains \cite{Obiedollah1}. Assuming that the CSI of all involved channels is perfect, the received signal of the $m$-th user in the $l$-th cluster can be given by
\begin{equation}\label{eq:1}
\begin{array}{l}
{\rm{y}}_{l,m}^{} = {G_t}{G_r}{{\bf{g}}_{l,m}}{\bf{\Theta}} {\bf{H}}(\sum\nolimits_{j = 1}^m {\sqrt {{p_{l,j}}} {s_{l,j}}}\vspace{0.5ex} \\
~~~~~~~~{\rm{                 }} + \sum\nolimits_{i \ne 1} {\sum\nolimits_{j = 1}^M {\sqrt {{p_{i,j}}} {s_{i,j}}} } ) + {n_{l,m}},\forall l,m,
\end{array}
\end{equation}
Similarly, the received signal of Eve is expressed as
\begin{equation}\label{eq:2}
{\rm{y}}_{l,m}^E = {G_t}{G_r}{{\bf{g}}_E}{\bf{\Theta H}}\sum\nolimits_{i = 1}^L {\sum\nolimits_{j = 1}^M {\sqrt {{p_{i,j}}} {s_{i,j}}} }  + {v_{l,m}},\forall l,m,
\end{equation}
where ${G_t}$ and ${G_r}$ represent the transmitting and receiving antenna gain, respectively, $ {\mathbf{H}} \in \mathbb{C}^{N_{\rm{{IRS}}} \times N_{\rm{{TX}}}} $ indicates the channel vector between BS and IRS, ${{\bf{g}}_{l,m}} \in {\mathbb{C} ^{1 \times {N_{{\rm{IRS}}}}}}$ denotes the equivalent channel from IRS to the $m$-th user in the $l$-th cluster, and 
${{\bf{g}}_E} \in {\mathbb{C}^{1 \times {N_{{\rm{IRS}}}}}}$ is  the channel vector from IRS to the Eve. The reflection phase shift matrix of IRS is defined by ${\bf{\Theta }} = {\rm{diag}}\left\{ {{ \bm{\theta} _1}, \ldots ,{ \bm{\theta} _n}} \right\}$, where $ \bm{\theta}_n   = {{\rm{e}}^{j{\theta  _n}}} $, ${\theta  _n} \in \left( {{\rm{0,2}}\pi } \right] $, and $ n \in\left\{ {1,\ldots,N_{\rm{{IRS}}}}\right\}$.
${s_{l,m}} = {\bf{F}}{{\bf{v}}_l}{x_{l,m}} $, ${x_{l,m}}$ is the transmission signal of the $m$-th user in the $l$-th cluster, 
${{\bf{v}}_l} \in \mathbb{C}{^{{N_{{\rm{RF}}}} \times 1}}$ denotes the digital precoding vector of the $l$-th cluster, and ${\bf{F}} \in\mathbb{C} {^{{N_{\rm{TX}}} \times {N_{{\rm{RF}}}}}}$ denotes the analog beamforming for all clusters, and $ {n_{l,m}}$ and ${v_{l,m}} \sim {\cal C}{\cal N}({\rm{0,}}\sigma ^2) $ are the additive white Gaussian noise (AWGN).
Moreover, THz is easily absorbed by water when propagating in the air, resulting in the scattering component being far lower than the line-of-sight (LoS) component \cite{Priebe}. Thus, we ignore scattering components of channels, and $\rm{\bf{H}}$ can be expressed as
\begin{equation}\label{eq:3}
{\bf{H}} = q(f,d)\overline {\bf{H}} ,
\end{equation}
where $q(f,d) = \frac{c}{{4\pi fd}}{e^{ - \frac{1}{2}\tau \left( f \right)d}}$ represents the path loss consisting of the free-space path loss and the molecular absorption loss in THz communications, $c$ is the speed of light, $f$ is operating frequency, $ \tau \left( f \right) $ represents the absorption coefficient of water molecules in the air, and $d$ is the distance from the BS to IRS. $\overline {\bf{H}}  = {\alpha _r}\left( \varphi  \right)\alpha _t^H\left( \vartheta  \right)$, where  $ {\alpha _r}\left( {{\varphi }} \right) $ and  $ {\alpha _t}\left( {{\vartheta}} \right) $ are the antenna array response vector of the transmitter and the receiver, respectively, which can be defined as 
\begin{equation}\label{eq:4}
{\alpha _r}\left( \varphi  \right){\rm{ = }}\frac{{\rm{1}}}{{\sqrt {{N_{{\rm{TX}}}}} }}\left[ {1,{e^{j\pi \varphi }},{e^{j2\pi \varphi }}, \cdots ,{e^{j\left( {{N_{{\rm{TX}}}} - 1} \right)\pi \varphi }}} \right],
\end{equation}
\begin{equation}\label{eq:5}
{\alpha _t}\left( \vartheta  \right) = \frac{{\rm{1}}}{{\sqrt {{N_{\rm{IRS}}}} }}\left[ {1,{e^{j\pi \vartheta }},{e^{j2\pi \vartheta }}, \cdots ,{e^{j\left( {{N_{{\rm{IRS}}}} - 1} \right)\pi \vartheta }}} \right].
\end{equation}
Here, $\varphi  = 2{d_0}f\sin ({\phi _t})/c$ and $\vartheta  = 2{d_0}f\sin ({\phi _r})/c$, where ${d_0}$ is the spacing between array elements, and
${\phi _i}\in \left[ { - {\pi  \mathord{\left/
			{\vphantom {\pi  2}} \right.
			\kern-\nulldelimiterspace} 2},{\pi  \mathord{\left/
			{\vphantom {\pi  2}} \right.
			\kern-\nulldelimiterspace} 2}} \right]$ , $i=[r,t]$, is the angle of departure (AOD) and the angle of arrival (AOA). Similarly, ${\bf{g}}_{l,m}^{}$
can be written as	
\begin{equation}\label{eq:6}
{{\bf{g}}_{l,m}} = q(f,{d_{l,m}}){\overline {\bf{g}} _{l,m}},
\end{equation}
where ${\overline {\bf{g}} _{l,m}} = \frac{1}{{\sqrt {{N_{\rm{IRS}}}} }}\left[ {1,{e^{j\pi {\varphi _{l,1,}}}},{e^{j2\pi {\varphi _{l,2}}}}, \cdots ,{e^{j\left( {{N_{{\rm{IRS}}}} - 1} \right)\pi {\varphi _{l,m}}}}} \right]$, and
$ q(f,{d_{l,m}}) = \frac{c}{{4\pi f{d_{l,m}}}}{e^{ - \frac{1}{2}\tau \left( f \right){d_{l,m}}}} $,
where ${d_{l,m}}$ denotes the distance from the IRS to the $m$-th user in the $l$-th cluster. 

Similarly, we can obtain
\begin{equation}\label{eq:7}
{{\bf{g}}_E} = q(f,{d_E}){\overline {\bf{g}} _E},
\end{equation}
where ${\overline {\bf{g}} _E} = \frac{1}{{\sqrt {{N_{\rm{IRS}}}} }}\left[ {1,{e^{j\pi {\varphi _,}}},{e^{j2\pi \varphi }}, \cdots ,{e^{j\left( {{N_{{\rm{IRS}}}} - 1} \right)\pi \varphi }}} \right]$,
$q(f,{d_{E}}) = \frac{c}{{4\pi f{d_{E}}}}{e^{ - \frac{1}{2}\tau \left( f \right){d_{E}}}}$, and $ {d_{E}} $ represents the distance between IRS and Eve.

For convenience, we set ${\beta _{l,m}} = \eta{G_t}{G_r}q(f,d)q(f,{d_{l,m}})$ and ${\beta _E} = \eta{G_t}{G_r}q(f,d)q(f,{d_E})$, where $\eta$ denotes the path compensation factor.
According to (3), (6) and (7), (1) and (2) can be rewriten as 
\begin{equation}\label{eq:8}
\begin{array}{l}
{y_{l,m}} = {{\bf{G}}_{l,m}}(\sum\nolimits_{j = {\rm{1}}}^m {\sqrt {{p_{l,j}}} {\bf{F}}{{\bf{v}}_l}{x_{l,j}}}\vspace{0.5ex}  \\
~~~~~~~~~~{\rm{            }} +\sum\nolimits_{i \ne l} {\sum\nolimits_{j = 1}^M {\sqrt {{p_{i,j}}} {\bf{F}}{{\bf{v}}_i}{x_{i,j}}} } ) + {n_{l,m}},\forall l,m,
\end{array}
\end{equation}
\begin{equation}\label{eq:9}
\hspace{-1.2cm}y_{l,m}^E = {{\bf{G}}_E}\sum\limits_{i = 1}^L {\sum\limits_{j = 1}^M {\sqrt {{p_{i,j}}} {\bf{F}}{{\bf{v}}_l}{x_{l,j}}} }  + {v_{l,m}}, \forall l,m,
\end{equation}
where ${{\bf{G}}_{l,m}}{\rm{ = }}{\beta _{l,m}}{\overline {\bf{g}} _{l,m}}{\bf{\Theta }}\overline {\bf{H}}$ with ${{\bf{G}}_{l,m}} \in \mathbb{C}{^{1 \times {N_{\rm{TX}}}}}$ , and ${{\bf{G}}_E} = {\beta _E} {\overline {\bf{g}} _E}{\bf{\Theta }}\overline {\bf{H}}$ 
with  ${\rm{ }}{{\bf{G}}_E} \in \mathbb{C}{^{1 \times {N_{\rm{TX}}}}}$.
\begin{figure}[tbp]
	\centering
	\subfigure[Fully-connected architecture]{\label{fig:a}\includegraphics[scale=0.6]{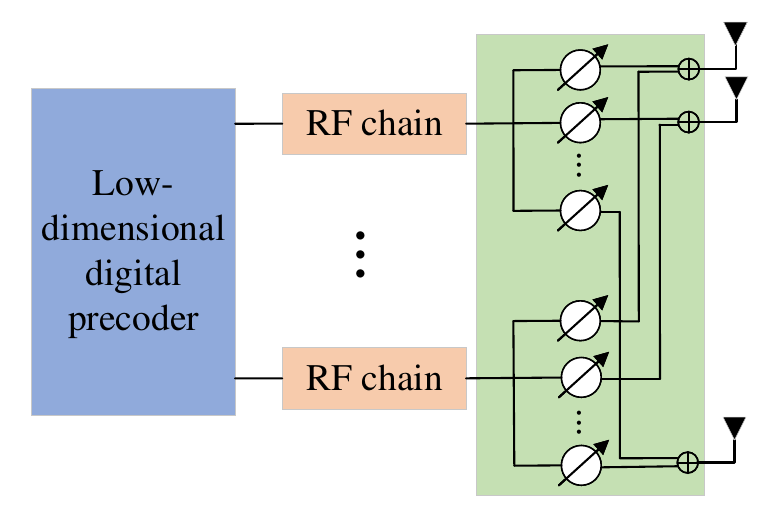}}\\
	\subfigure[Sub-connected HP architecture]{\label{fig:b}\includegraphics[scale=0.6]{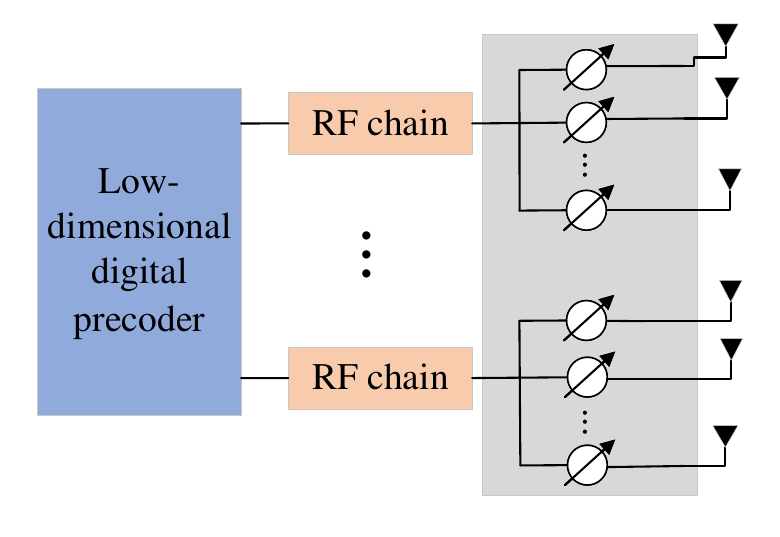}}\\	
	\caption{Sparse RF chain antenna structures.}
\end{figure}
\subsection{Analog Beamforming}
Fig. 2 shows the architecture of the studied THz MIMO system. As shown in Fig. 2 (a), there are ${N_{\rm{TX}}} \times {N_{\rm{RF}}} $  phase shifters in the FC HP architecture. ${\bf{F}} \in\mathbb{C} {^{{N_{\rm{TX}}} \times {N_{{\rm{RF}}}}}}$ represents the analog precoding matrix implemented by equal power dividers and phase shifters. Then, ${{\bf{F}}^{(full)}}$ can be given by
\begin{equation}\label{eq:10}
{{\bf{F}}^{(full)}} = \left[ {{\bf{f}}_1^{(full)},{\bf{f}}_2^{(full)}{\rm{,}}{\bf{f}}_{\rm{q}}^{(full)}{\rm{,}} \ldots {\rm{,}}{\bf{f}}_{{N_{\rm{RF}}}}^{(full)}} \right],
\end{equation}
where ${{\bf{f}}_q} \in \mathbb{C} {^{{N_{{\rm{TX}}}} \times 1}}$, $\forall q = \left\{ {1,2, \ldots ,{N_{\rm{RF}}}} \right\}$ is the analog precoding vector associated with the $q$-th RF chain.
Moreover, Fig. 2 (b) is the SC HP architecture. Without loss of generality, we assume that ${N_{\rm{SF}}} = {{{N_{{\rm{TX}}}}} \mathord{\left/
		{\vphantom {{{N_{{\rm{TX}}}}} {{N_{\rm{RF}}}}}} \right.
		\kern-\nulldelimiterspace} {{N_{\rm{RF}}}}}$ is an integer number, and each RF chain is connected to ${N_{\rm{SF}}}$  phase shifters \cite{Haijun1}. For the SC architecture, the precoding matrix ${{\bf{F}}^{(sub)}}$ is
\begin{equation}\label{eq:11}
{{\bf{F}}^{(sub)}} = \left[ \begin{array}{l}
{\rm{ f}}_1^{(sub)}{\rm{  \hspace{0.3cm}  }}0\hspace{0.3cm}{\rm{   }} \cdots \hspace{0.3cm}{\rm{   0 }}\\
{\rm{ 0 \hspace{0.9cm}      f}}_2^{(sub)}\hspace{0.6cm}{\rm{       0 }}\\
{\rm{ }}\hspace{0.05cm}  \vdots \hspace{0.9cm} {\rm{              }} \ddots\hspace{0.95cm}  {\rm{   }} \vdots \\
{\rm{ 0 \hspace{0.9cm}       0    }} \hspace{0.3cm}\cdots \hspace{0.1cm}{\rm{  f}}_{{N_{\rm{RF}}}}^{(sub)}
\end{array} \right].
\end{equation}
Compared with the FC structure, the SC structure reduces the hardware cost and power consumption at the expense of reduced array gain\cite{Mengyuan}.

For practical implementation, we consider a $B$ bit quantization phase shifter \cite{Linglong1}. The FC analog non-zero element precoding matrix ${{\bf{F}}^{(full)}}$ can be given by
\begin{equation}\label{eq:12}
{{\bf{F}}^{(full)}}={{\rm{1}} \mathord{\left/
		{\vphantom {{\rm{1}} {\sqrt {{N_{\rm{TX}}}} }}} \right.
		\kern-\nulldelimiterspace} {\sqrt {{N_{\rm{TX}}}} }}\left\{ {{e^{j\frac{{2\pi \beta }}{{{2^B}}}}}:\beta = 0,1, \ldots ,{2^B} - 1} \right\}.
\end{equation}
where ${\bf{f}}_l^{(full)}\left( q \right) = \frac{1}{{\sqrt {{N_{{\rm{BS}}}}} }}{e^{j\frac{{2\pi \widehat \beta}}{{{2^B}}}}}$.
Similarly, the SC analog precoding matrix ${{\bf{F}}^{(sub)}}$ can be written as
\begin{equation}\label{eq:13}
{{\bf{F}}^{(sub)}}={{\rm{1}} \mathord{\left/
		{\vphantom {{\rm{1}} {\sqrt {{N_{\rm{SF}}}} }}} \right.
		\kern-\nulldelimiterspace} {\sqrt {{N_{\rm{SF}}}} }}\left\{ {{e^{j\frac{{2\pi \beta}}{{{2^B}}}}}:\beta = 0,1, \ldots ,{2^B} - 1} \right\}.
\end{equation}
In (13), ${\bf{f}}_l^{(sub)}\left( q \right) = \frac{1}{{\sqrt {{N_{{\rm{SF}}}}} }}{e^{j\frac{{2\pi \widehat \beta}}{{{2^B}}}}}$, $(q = (L-1){N_{{\rm{SF}}}}+1,(L-1){N_{{\rm{SF}}}}+2, \ldots ,L{N_{{\rm{SF}}}})$ represents the $q$-th element of SC analog precoding vector,
where $\widehat \beta = \arg \min \left| {angle\left( {{\bf{g}}\left( l \right)} \right) - \frac{{2\pi \beta}}{{{2^B}}}} \right|$, and ${\bf{g}}\left( l \right)$ is the channel state information for $l$-th cluster head.

\subsection{Digital Precoding}
Since each cluster contains multiple users, it will cause inter-cluster interference \cite{Wanming2}, we adopt the low-complexity ZF method to design the digital precoding and eliminate interference among cluster head users \cite{Bichai1}.
Assuming that equivalent channel gain of all cluster head users is
\begin{equation}\label{eq:14}
{\bf{G}} = \left[ {{{\bf{g}}_{1,m}},{{\bf{g}}_{2,m}}{\rm{,}} \ldots {\rm{,}}{{\bf{g}}_{L,m}}} \right].
\end{equation}
Then, we can generate the ZF precoding matrix as
\begin{equation}\label{eq:15}
\overline {\bf{V}}  = \left[ {{{\overline {\bf{v}} }_1},{{\overline {\bf{v}} }_2}{\rm{,}} \ldots {\rm{,}}{{\overline {\bf{v}} }_L}} \right] = {\bf{G}}{\left( {{{\bf{G}}^H}{\bf{G}}} \right)^{ - 1}},
\end{equation}
After normalization, the digital precoding vector of the $l$-th beam can be written as ${{\bf{v}}_l} = \frac{{{{\overline {\bf{v}} }_l}}}{{{{\left\| {{\bf{F}}{{\overline {\bf{v}} }_l}} \right\|}_2}}}$.

\subsection{Secrecy Rate}
Upon performing the HP of two antenna structures, the SINR of the $m$-th user of the $l$-th cluster and Eve can be expressed as  
\begin{equation}\label{eq:16}
~~{\gamma _{l,m}} = \frac{{{{\left| {{{\bf{G}}_{l,m}}{\bf{F}}{{\bf{v}}_l}} \right|}^2}{p_{l,m}}}}{{{{\left| {{{\bf{G}}_{l,m}}{\bf{F}}{{\bf{v}}_l}} \right|}^2}\sum\limits_j^{m - 1} {{p_{l,j}}}  + \sum\limits_{i \ne l} {{{\left| {{{\bf{G}}_{l,m}}{\bf{F}}{{\bf{v}}_i}} \right|}^2}\sum\limits_{j = 1}^M {{p_{i,j}}} }  + {\sigma ^2}}},
\end{equation}
\begin{equation}\label{eq:17}
\gamma _{l,m}^E = \frac{{{{\left| {{{\bf{G}}_E}{\bf{F}}{{\bf{v}}_l}} \right|}^2}{p_{l,m}}}}{{{{\left| {{{\bf{G}}_E}{\bf{F}}{{\bf{v}}_l}} \right|}^2}\sum\limits_{j \ne m} {{p_{l,j}}}  + \sum\limits_{i \ne l} {{{\left| {{{\bf{G}}_E}{\bf{F}}{{\bf{v}}_i}} \right|}^2}\sum\limits_{j = 1}^M {{p_{i,j}}} }  + {\sigma ^2}}},~
\end{equation}
Then, the secrecy rate of the $m$-th user of the $l$-th cluster can be expressed as
\begin{equation}\label{eq:18}
\begin{array}{l}
R_{l,m}^{sec} =\hspace{0.1cm}{\log _2}\left( {1 + {\gamma _{l,m}}} \right) - {\log _2}( {1 + \gamma _{l,m}^E} ).
\end{array}
\end{equation}

\subsection{Problem Formulation}
Our goal is maximizing the sum secrecy rate by jointly optimize the power alloction and the phase-shift matrix, the optimization problem is formulated as 
\begin{subequations}\label{eq:19}
	\begin{align}
	\mathop {\max }\limits_{{p_{l,m}},{\bf{\Theta }}} {\rm{     }}	&~\sum\limits_{l = 1}^L {\sum\limits_{m = 1}^M {R_{l,m}^{sec}} }  \\
	\;\;{\rm{ s}}{\rm{.t}}{\rm{.~}}
	&~\sum\limits_{l = 1}^L {\sum\limits_{m = 1}^M {{p_{l,m}} \le {P_T}} } , \\
	&~{\log _2}\left( {1 + {\gamma _{l,m}}} \right) \ge R_{l,m}^{\min }, \\
	&\left| {{\theta _n}} \right| = 1,\forall n \in {N_{\rm{IRS}}}.
	\end{align}
\end{subequations}
where ${{P_T}}$ stands for the transmit power of the BS, (19b)  represents the power constraint of the system, (19c) indicates the rate requirement of each user, 
and (19d) represents the unit modulus constraint of each IRS element. Problem (19) is a non-convex optimization problem and intractable to solve due to the coupled variables. 
To overcome this issue, we decomposed (19) into two sub-problems. For each sub-problem, we only focus on some specific variables when the others are fixed, which will be presented in the section III.
\vspace{3ex}

\section{Joint Optimize the Power Alloction and the Phase Shifts Matrix}
In this section,  an efficient AO algorithm is utilized to solve problem (19). Specifically, we firstly decouple the origan problem into two sub-problems. Then, the first-order Taylor expansion and SDR are leveraged to transform each sub-problem into a convex counterpart. 
Finally, the local optimal solution can be obtained through the AO algorithm.
\subsection{Optimal the Power Allocation }
To overcome the above issue, we assume that the order of the effective channel gain satisfies ${\left| {{{\overline {\bf{G}} }_{l,1}}} \right|^2} \ge {\left| {{{\overline {\bf{G}} }_{l,2}}} \right|^2} \ge  \cdots  \ge {\left| {{{\overline {\bf{G}} }_{l,m}}} \right|^2}$, i.e. the SIC decoding sequence of NOMA users. Moreover, defining ${{\bf{G}}_{l,m}}{\bf{F}}  \buildrel \Delta \over =  {\overline {\bf{G}} _{l,m}}$ and ${\bf{G}}_{l,m}^E{\bf{F}} \buildrel \Delta \over = \overline {\bf{G}} _{l,m}^E$, the (18) can be rewritten as 
\begin{equation}\label{eq:20}
{\overline {R}}_{l,m}^{sec} = {\rm{lo}}{{\rm{g}}_2}\left( {{A_1}} \right) - {\log _2}\left( {{B_1}} \right) - {\log _2}\left( {{C_1}} \right) + {\log _2}\left( {{D_1}} \right),
\end{equation}
where ${{B_1}}$ can be given by 
\begin{equation}\label{eq:21}
\begin{array}{l}
\hspace{-1.1ex}{B_1}\hspace{-0.5ex} = \hspace{-0.5ex}{\left| {{{\overline {\bf{G}} }_{l,m}}{{\bf{v}}_l}} \right|^2}\sum\limits_j^{m - 1} {{p_{l,j}}} \hspace{-0.5ex} +\hspace{-0.5ex} \sum\limits_{i \ne l} {{{\left| {{{\overline {\bf{G}} }_{l,m}}{{\bf{v}}_i}} \right|}^2}\sum\limits_{j = 1}^M {{p_{i,j}}} }  + {\sigma ^2},
\end{array}
\end{equation}
Then, ${A_1}$ can be expressed as
\begin{equation}\label{eq:22}
\begin{array}{l}
\hspace{-0.15cm}{A_1}{\rm{ = }}{B_1} + {\left| {{{\overline {\bf{G}} }_{l,m}}{{\bf{v}}_l}} \right|^2}{p_{l,m}}\\
~~{\rm{  }}\mathop  ={\left| {{{\overline {\bf{G}} }_{l,m}}{{\bf{v}}_l}} \right|^2}\sum\limits_j^{m} {{p_{l,j}}}  + \sum\limits_{i \ne l} {{{\left| {{{\overline {\bf{G}} }_{l,m}}{{\bf{v}}_i}} \right|}^2}\sum\limits_{j = 1}^M {{p_{i,j}}} }  + {\sigma ^2},
\end{array}
\end{equation}
Similarly, ${D_1}$ can be given by
\begin{equation}\label{23}
\begin{array}{l}
\hspace{-0.414cm}{D_1}= {\left| {\overline {\bf{G}} _{l,m}^E{{\bf{v}}_l}} \right|^2}\sum\limits_{j \ne m} {{p_{l,j}}}  + \sum\limits_{i \ne l} {{{\left| {\overline {\bf{G}} _{l,m}^E{{\bf{v}}_i}} \right|}^2}\sum\limits_{j = 1}^M {{p_{i,j}}} }  + {\sigma ^2},~\\
\end{array}
\end{equation}
Further, we can express ${C}_1$ as
\begin{equation}\label{24}
\begin{array}{l}
{{C}_1} = {D_1} + {\left| {\overline {\bf{G}} _{l,m}^E{{\bf{v}}_l}} \right|^2}{p_{l,m}}
{\rm{      = }}\sum\limits_{i = l}^L {{{\left| {\overline {\bf{G}} _{l,m}^E{{\bf{v}}_l}} \right|}^2}} \sum\limits_{j = 1}^M {{p_{i,j}}}  + {\sigma ^2}.~~~~~~
\end{array}
\end{equation}

Due to the existence of logarithmic term $- {\log _2}\left( B_1 \right) - {\log _2}\left( C_1 \right)$, problem (20) is still non-convex \cite{Zhaofei}. To
deal with this problem, we apply the first-order Taylor expansion on the point 
${{{\overline p }_{l,m}}}$\cite{Con}. For convenience, we set ${\overline {B}_1} = {\left| {{{\overline {\bf{G}} }_{l,m}}{{\bf{v}}_l}} \right|^2}\sum\limits_j^{m - 1} {{{\overline p }_{l,j}}}  + \sum\limits_{i \ne l} {{{\left| {{{\overline {\bf{G}} }_{l,m}}{{\bf{v}}_i}} \right|}^2}\sum\limits_{j = 1}^M {{{\overline p }_{i,j}}} }  + {\sigma ^2}$, and ${\overline {C}_1}{\rm{      = }}\sum\limits_{i = l}^L {{{\left| {\overline {\bf{G}} _{l,m}^E{{\bf{v}}_l}} \right|}^2}} \sum\limits_{j = 1}^M {{\overline p_{i,j}}}  + {\sigma ^2}$. Then, we can transform 
$- {\log _2}\left( B_1 \right)$ and $- {\log _2}\left( C_1 \right)$
into convex forms, respectively, as
\begin{equation}\label{25}
\hspace{-0.72cm}{\rm{lo}}{{\rm{g}}_2}\left( {{B_1}} \right) \le  {\log _2}({\overline B _1}) + \frac{{ {\overline {\bf{G}} _{l,m}^{}{\bf{\Omega}} \overline {\bf{G}} _{l,m}^H} }}{{{{\overline B }_1}\ln 2}},
\end{equation}
\begin{equation}\label{26}
\hspace{-0.35cm}{\rm{lo}}{{\rm{g}}_2}\left( {{C_1}} \right) \le  {\log _2}({\overline C _1}) + \frac{{ {\overline {\bf{G}} _{l,m}^E{\bf{\Psi }}\overline {\bf{G}} {{_{l,m}^E}^H}} }}{{{{\overline C }_1}\ln 2}},
\end{equation}
where ${\bf{\Omega} }  = {\left| {{{\bf{v}}_l}} \right|^2}\sum\limits_{j = 1}^{m - 1} {\left( {{p_{i,j}} - {{\overline p}_{i,j}}} \right)}  + \sum\limits_{i \ne l}^L {{{\left| {{{\bf{v}}_i}} \right|}^2}\sum\limits_{j = 1}^M {\left( {{p_{i,j}} - {{\overline p}_{i,j}}} \right)} } $ and ${\bf{\Psi }} = \sum\limits_{i = 1}^L {{{\left| {{{\bf{v}}_i}} \right|}^2}\sum\limits_{j = 1}^M {\left( {{p_{i,j}} - {{\overline p }_{l,j}}} \right)} } $. Obviously, (25) and (26) are both linear. According to the above derivation, the constraint (19c) can be transformed as 
\begin{equation}\label{27}
{p_{l,m}}{\left| {{{\bf{v}}_l}} \right|^2}{\overline {\bf{G}} _{l,m}\overline {\bf{G}} _{l,m}^H}  - \left( {{2^{R_{l,m}^{\min }}} - 1} \right){{{B}}_1} \ge 0.
\end{equation}

Finally, the problem (19) can be recast as 
\begin{subequations}\label{eq:28}
	\begin{align}
	\mathop {\max }\limits_{{p_{l,m}},{\bf{\Theta }}} {\rm{     }}	&~\sum\limits_{l = 1}^L {\sum\limits_{m = 1}^M {R_{l,m}^{sec}} }  \\
	\rm{s.t.}     &~(19\rm{b}),(27).
	\end{align}
\end{subequations}
With the given ${\bf{\Theta }}$, the problem (28) can be computed by applying convex problem solvers such as CVX toolbox \cite{CVX}. The algorithm of optimizing power allocation is summarized to algorithm 1.
However, since the approximation algorithm may cause the problem to diverge. Then, we assume that $p_{l,m}^{({t_1})}$ is the optimal solution  for the ${t_1}$ th iteration, and the convergence of algorithm 1 is proved in the following theorem.

\begin{algorithm}
    \caption{\hspace{-0.5ex}{\bf{{:}}} The optimization algorithm for the power allocation.}
	\label{Algorithm2}
	\begin{algorithmic}[1]
		\STATE Initialize the iteration number
			$t_1 = 0$ and give a feasible solution $p_{l,m}^{({t_0})}$ for the problem (28).\\
		\REPEAT
		\STATE Leveraging the $p_{l,m}^{({t_0})}$ to obtain a optimal solution $p_{l,m}^{({t_1})}$ for problem (28).\\		
			\STATE Update the feasible solution $p_{l,m}^{({t_1})} \leftarrow p_{l,m}^{({t_1} - 1)}$.\\		
\STATE Update the iteration number ${t_1} \leftarrow {t_1} - 1$.
		\UNTIL $\left| {{\rm{lo}}{{\rm{g}}_2}\left( {B_1^{({t_1}  )}} \right) - {\rm{lo}}{{\rm{g}}_2}\left( {B_1^{({t_1}-1)}} \right)} \right| \le \xi $ and $\left| {{\rm{lo}}{{\rm{g}}_2}\left( {C_1^{({t_1})}} \right) - {\rm{lo}}{{\rm{g}}_2}\left( {C_1^{({t_1-1})}} \right)} \right| \le \xi $.
	\end{algorithmic}
\end{algorithm}

{\bf{{{Theorem 1}}}}: Defining $ f({t_1})$ as the objective value of the ${t_1}$-th iteration, we have $f({t_1} ) \ge f({t_1}-1)$. Moreover, defining $\left\| {{\rm{\Delta }}\ell } \right\| \buildrel \Delta \over = f({t_1} ) - f({t_1-1})$, the value of $\left\| {{\rm{\Delta }}\ell } \right\|$ infinitely close to 0 with the ${t_1}$ increases, i.e., the algorthim converges.

{{\emph{{Proof}}}}: \rm{See Appendix A.}
$\hfill\blacksquare$

\subsection{Solving the Phase Shifts of IRS }
With the obtained power alloction ${{p_{l,m}}}$, the original problem (19) can be transformed a feasibility check problem.
By denoting ${\bm{\theta}} {\rm{ = }}{\left[ {{\theta _{\rm{1}}}, \ldots ,{\theta _{N_{\rm{IRS}}}}} \right]^T}$, the equivalent forms of the cascaded channels can be respectively expressed as
\begin{equation}\label{29}
{\overline {\bf{g}} _{l,m}}{\bm{\Theta }}\overline {\bf{H}}  = {\theta ^H}{\rm{diag}}\left({\overline {\bf{g}} _{l,m}}\right)\overline {\bf{H}}  = { \bm{\theta} ^H}{\widetilde {\bf{G}}_{l,m}},
\end{equation}
\vspace{-1.5ex}
\begin{equation}\label{30}
\overline {\bf{g}} _{l,m}^E{\bm{\Theta }}\overline {\bf{H}}  = {\theta ^H}{\rm{diag}}\left(\overline {\bf{g}} _{l,m}^E\right)\overline {\bf{H}}  = {  \bm{\theta}^H}\widetilde {\bf{G}}_{l,m}^E,
\end{equation}

Then, we reconsider the phase shift matrix ${\bf{\Theta }}$ and transform the secrecy rate into the following form
\begin{equation}\label{31} 
{\widehat {R}}_{l,m}^{sec} = {\rm{lo}}{{\rm{g}}_2}\left( {{A_2}} \right) + {\log _2}\left( {{D_2}} \right) - {\log _2}\left( {{B_2}} \right) - {\log _2}\left( {{C_2}} \right).
\end{equation}

For the non-convex objective function, we denote ${\bm{\Phi }}\buildrel \Delta \over =\bm{\theta} {\bm{\theta} }^H$, ${\bf{V}} \buildrel \Delta \over = {\bf{v}}{{\bf{v}}^H}$,
${\widehat {\bf{G}}_{l,m}} \buildrel \Delta \over = {\beta _{l,m}}{\widetilde {\bf{G}} _{l,m}}{\bf{F}}$ 
and $\widehat {\bf{G}}_{l,m}^E \buildrel \Delta \over = {\beta _E}\widetilde {\bf{G}} _{l,m}^E{\bf{F}}$. 
As such, the $B_2$ in (31) can be expressed as 
\begin{equation}\label{32}
\begin{array}{l}
\hspace{-0.2cm}{B_2} = {\left| {{ \bm{\theta} ^H}{{\widehat {\bf{G}}}_{l,m}}{{\bf{v}}_l}} \right|^2}\sum\limits_j^{m - 1} {{p_{l,j}}}  + \sum\limits_{i \ne l} {{{\left| {{ \bm{\theta} ^H}{{\widehat {\bf{G}}}_{l,m}}{{\bf{v}}_i}} \right|}^2}\sum\limits_{j = 1}^M {{p_{i,j}}} }  + {\sigma ^2}\\
\hspace{0.24cm}{\rm{    }}\mathop  = \limits^{(b)} {\rm{Tr   }}( {{{\widehat {\bf{G}}}_{l,m}}( {{{\bf{V}}_l}\sum\limits_{j{\rm{ = 1}}}^{m - 1} {{p_{l,j}}}  + \sum\limits_{i \ne l} {{{\bf{V}}_i}\sum\limits_{j = 1}^M {{p_{i,j}}} } } )\widehat {\bf{G}}_{l,m}^H {\bm{\Phi }}} ) + {\sigma ^2},
\end{array}
\end{equation}
where ${\widehat {\bf{G}}_{l,m}} \buildrel \Delta \over = {\beta _{l,m}}{\widetilde {\bf{G}} _{l,m}}{\bf{F}}$, 
and $(b)$ is obtained by invoking the identity ${{\rm{\bf{a}}}^H}{\bf{T}}{\rm{\bf{a} = Tr}}\left( {{\bf{T}}{\rm{\bf{a}}}{{\rm{\bf{a}}}^H}} \right)$. Then, $A_2$ can be written as
\vspace{-2ex}
\begin{equation}\label{33}
\vspace{-0.5ex}
\begin{array}{l}
\hspace{-0.3cm}{{{A}}_2}{\rm{ = }}{B_2} \hspace{-0.5ex} +\hspace{-0.5ex}  {\left| {{ \bm{\theta} ^H}{{\widehat {\bf{G}}}_{l,m}}} \right|^2}{{\bf{V}}_l}{p_{l,m}}\\
~\hspace{-0.1cm}{\rm{    }}\mathop  = \limits^{(b)}{\rm{Tr   }}( {{{\widehat {\bf{G}}}_{l,m}}( {{{\bf{V}}_l}\sum\limits_{j{\rm{ = 1}}}^m {{p_{l,j}}}  \hspace{-0.5ex} +\hspace{-0.5ex}  \sum\limits_{i \ne l} {{{\bf{V}}_i}\sum\limits_{j = 1}^M {{p_{i,j}}} } } )\widehat {\bf{G}}_{l,m}^H {\bm{\Phi }}} ) \hspace{-0.5ex} +\hspace{-0.5ex}  {\sigma ^2},
\end{array}
\end{equation}
According to the previous method, $D_2$ can be rewritten  as 
\begin{equation}\label{34}
\begin{array}{l}
\hspace{-0.35cm}{{{D}}_2} = {\left| {{ \bm{\theta} ^H}\widehat {\bf{G}}_{l,m}^E{{\bf{v}}_l}} \right|^2}\sum\limits_{j \ne m} {{p_{l,j}}} \hspace{-0.5ex} +\hspace{-0.5ex} \sum\limits_{i \ne l} {{{\left| {{ \bm{\theta} ^H}\widehat {\bf{G}}_{l,m}^E{{\bf{v}}_i}} \right|}^2}\sum\limits_{j = 1}^M {{p_{i,j}}} }  \hspace{-0.5ex} +\hspace{-0.5ex} {\sigma ^2}\\
\hspace{0.1cm}{\rm{    }}\mathop  = \limits^{(b)}{\rm{Tr   }}( {\widehat {\bf{G}}_{l,m}^E( {{{\bf{V}}_l}\sum\limits_{j \ne m} {{p_{l,j}}}  \hspace{-0.5ex} +\hspace{-0.5ex} \sum\limits_{i \ne l} {{{\bf{V}}_i}\sum\limits_{j = 1}^M {{p_{i,j}}} } } )\widehat {\bf{G}}{{_{l,m}^E}^H} {\bm{\Phi }}} ){\rm{ + }}{\sigma ^2},
\end{array}
\end{equation}
and $C_2$ can be expressed as
\begin{equation}\label{35}
\begin{array}{l}
\hspace{-0.3cm}{{{C}}_2} = {{{D}}_2} \hspace{-0.5ex} +\hspace{-0.5ex}  {\left| {\left| {{\bm{\theta}}^H\widehat {\bf{G}}_{l,m}^E{{\bf{v}}_l}} \right|} \right|^2}{p_{l,m}}\\
\hspace{0.1cm}{\rm{    }}\mathop {\rm{ = }}\limits^{(b)} {\rm{Tr}}( {\widehat {\bf{G}}_{l,m}^E( {{{\bf{V}}_l}\sum\limits_{j = 1}^m {{p_{l,j}}}  \hspace{-0.5ex} +\hspace{-0.5ex}  \sum\limits_{i = l} {{{\bf{V}}_i}\sum\limits_{j = 1}^M {{p_{i,j}}} } } )\widehat {\bf{G}}{{_{l,m}^E}^H} {\bm{\Phi }}} ) \hspace{-0.5ex} +\hspace{-0.5ex}  {\sigma ^2}.
\end{array}
\vspace{-0.5ex}
\end{equation}

Similar approximation method previously adopted can be utilized for the non-convex item $- {\log _2}\left( B_2 \right) - {\log _2}\left( C_2 \right)$ in (31), the first-order Taylor expansion is leveraged on the point $\widehat {\bf{\Phi }}$ \cite{Zhaofei}. For implicity, we set 
\begin{equation}\label{36}
\vspace{-0.5ex}
{\widehat B_2} = {\rm{Tr}}( {{{\widehat {\bf{G}}}_{l,m}}( {{{\bf{V}}_l}\sum\limits_{j{\rm{ = 1}}}^{m - 1} {{p_{l,j}}}  \hspace{-0.5ex} +\hspace{-0.5ex} \sum\limits_{i \ne l} {{{\bf{V}}_i}\sum\limits_{j = 1}^M {{p_{i,j}}} } } )\widehat {\bf{G}}_{l,m}^H\widehat {\bf{\Phi }}} ) \hspace{-0.5ex} +\hspace{-0.5ex} {\sigma ^2},
\end{equation}
\vspace{-2ex}
\begin{equation}\label{37}
{\widehat {\rm{C}}_2} = {\rm{Tr}}( {\widehat {\bf{G}}_{l,m}^E} ( {{{\bf{V}}_l}\sum\limits_{j = 1}^m {{p_{l,j}} \hspace{-0.5ex} +\hspace{-0.5ex}  } } 
{\sum\limits_{i = l} {{{\bf{V}}_i}\sum\limits_{j = 1}^M {{p_{i,j}}} } } ) {\widehat {\bf{G}}{{_{l,m}^E}^H} \widehat {\bf{\Phi }}} ) \hspace{-0.5ex} +\hspace{-0.5ex}  {\sigma ^2}
\end{equation}
Then,
$- {\log _2}\left( B_2 \right)$ and $- {\log _2}\left( C_2 \right)$ can be transformed 
into convex forms, respectively, as
\begin{equation}\label{38}
{\rm{lo}}{{\rm{g}}_2}( {{{{B}}_2}} ) \le  {\log _2}({\widehat {{B}}_2}) + \frac{{{\rm{Tr}}( {{{\widehat {\bf{G}}}_{l,m}}\widehat{\bm{\Omega}} \overline {\bf{G}} _{l,m}^H( {{\bf{\Phi }} - \widehat {\bf{\Phi }}} )} )}}{{{{\widehat {{B}}}_2}\ln 2}},
\end{equation}
\begin{equation}\label{39}
~~{\rm{lo}}{{\rm{g}}_2}\left( {{C_2}} \right) \le  {\log _2}({\widehat C_2}) + \frac{{{\rm{Tr}}( {\widehat {\bf{G}}_{l,m}^E\widehat {\bf{\Psi }}\widehat {\bf{G}}{{_{l,m}^E}^H}( {{\bf{\Phi }} - \widehat {\bf{\Phi }} } )} )}}{{{{\widehat C}_2}\ln 2}}.
\end{equation}
where ${\bf{\widehat \Omega}} {\rm{ = }}{{\bf{V}}_l}\sum\limits_{j = 1}^m {{p_{l,j}}}  + \sum\limits_{i \ne l} {{{\bf{V}}_i}\sum\limits_{j = 1}^M {{p_{i,j}}} } $ and $\widehat {\bf{\Psi }} = {{\bf{V}}_l}\sum\limits_{j{\rm{ = 1}}}^{m - 1} {{p_{l,j}}}  + \sum\limits_{i \ne l} {{{\bf{V}}_i}\sum\limits_{j = 1}^M {{p_{i,j}}} } $.
Then, we plug (38) and (39) into constraint
(31), and transform constraint (19c) into convex form as
\begin{equation}\label{40}
{p_{l,m}}{\rm{Tr}}\left( {{{\widehat {\bf{G}}}_{l,m}}{{\bf{V}}_l}\widehat {\bf{G}}_{l,m}^H{\bf{\Phi }}} \right) - \left( {{2^{R_{l,m}^{\min }}} - 1} \right){{{B}}_2} \ge 0.
\end{equation}

As a result, the passive beamforming optimized can be recast as
\begin{subequations}\label{eq:41}
	\begin{align}
	\mathop {\max }\limits_{\bf{\Phi }} {\rm{   }}	&~\sum\limits_{l = 1}^L {\sum\limits_{m = 1}^M {R_{l,m}^{sec}} } \\
	\;\;{\rm{ s}}{\rm{.t}}{\rm{.~}}
	&~{{\bf{\Phi }}_{n,n}} = 1,\forall n \in {N_{{\rm{IRS}}}}, \\
	&~\rm{rank}({\bf{\Phi }}) = 1,{\bf{\Phi }}\succeq 0, \\
	&~(36).  
	\end{align}
\end{subequations}
However, due to the constraint $\rm{rank}({\bf{\Phi }}) = 1$,
it is difficult to directly to address it by CVX. Therefore, we neglect the rank-one constraint via SDR \cite{Chu4}, and the problem (41) can be rewritten as
\begin{subequations}\label{eq:42}
	\begin{align}
	\mathop {\max }\limits_{\bf{\Phi }} {\rm{   }}	&~\sum\limits_{l = 1}^L {\sum\limits_{m = 1}^M {R_{l,m}^{sec}} } \\
	\;\;{\rm{ s}}{\rm{.t}}{\rm{.~}}
	&~{\bf{\Phi }}\succeq 0,\\
	&~\rm{(36),(37b)}.
	\end{align}
\end{subequations}

\begin{algorithm}
	\caption{\hspace{-0.5ex}{\bf{{:}}} The overall AO algorithm }
	\label{Algorithm2}
	\begin{algorithmic}[1]
		\STATE \textbf{Initialize:}{ reflection matrix ${\bf{\Theta}}^{0}$, iteration number
			$t = 1$ and maximum iteration number $T$}.
		\STATE Obtain the analog beamforming ${{\bf{F}}^{(full)}}$ and ${{\bf{F}}^{(sub)}}$. Then, the digital beamforming $\bf{V}$ is designed by the ZF precoding. 
		\REPEAT
		\REPEAT
		\STATE Given $\bf{\Theta}$, solve the problem (28) and obtain the solution $p_{l,m}^{(t_1)}$.
		
		\STATE Update {$t_1 \leftarrow t_1+1$}.
		
		\STATE Update {$p_{l,m}^{(t_1)} \leftarrow {p_{l,m}^{(t_1)}}^*$}.
		\UNTIL{$t_1=T_1$ or Convergence;}\\
		\REPEAT
		\STATE with optimized ${p_{l,m}}^*$, the solution  ${\bf{\Phi}}^{(t_2)}$ of the problem (42) can be obtained.
		
		\STATE Update {$t_2 \leftarrow t_2+1$}.
		
		\STATE Update {${\bf{\Phi}}^{(t_2)} \leftarrow {{\bf{\Phi}}^{(t_2)}}^*$}.
		\UNTIL{$t_2=T_2$ or Convergence;}\\
		\UNTIL{$t=T$ or Convergence;}\\
		Maximize the achievable sum rate via the obtained  ${p_{l,m}}^*$ and ${{\bf{\Phi}}^*}$.
	\end{algorithmic}
\end{algorithm}

It is clear that problem (42) is a standard SDP, which can be efficiently solved by applying the CVX toolbox \cite{CVX}. 
Generally speaking, after ignoring the $\rm{rank}({\bf{\Phi }}) = 1$, the solution calculated with eigenvalue decomposition  for $\bf{\Phi }$ does not necessarily satisfy rank one.
Therefore, the optimal result of the problem (42) only serves as an upper bound of (41) \cite{Qing}. In order to construct a rank one solution from the optimal higher-order solution of problem (42),  singular value decomposition is performed on ${\bf{\Phi }} = {\bm{U}}{\bm{\Sigma}} {\bm{U}^H}$, where $\bm{U}{\rm{ = [}}{e_1}, \cdots ,{e_{N }}{\rm{]}}$ and $\bm{\Sigma}  = {{\rm{diag}(}}{\lambda _1}, \cdots ,{\lambda _{N }})$ are a unitary matrix and a diagonal matrix respectively. 
Then, the suboptimal solution ${\bm{\theta ^*}} = \bm{U}{\bm{\Sigma} ^{1/2}}\bm{r} $   of (41) is obtained, 
where $ {\bm{r}} \in \mathbb{C}  {^{\left( {N } \right) \times 1}}$ is a random vector, each element of which is a Gaussian random variable with zero mean and unit variance, i.e., $ \bm{r} \sim {\cal C}{\cal N}({\rm{0,}}{\bm{I}_{N}})$.
Using the independently generated Gaussian random vector $\bm{r}$, the objective value of (42) is approximated to the maximum value for all the optimal $\bm{\theta}$ \cite{Qing2}. Finally, we can obtain $\bm{\theta}$ by
\begin{equation}\label{43}
{\theta _n} = {e^{j\angle \left( {\left[ {\frac{{{{{\bm{\theta}}} _n}}}{{{{{\bm{\theta}}} _{n + 1}}}}} \right]} \right)}},n = 1,2, \cdots ,N.
\end{equation}
In this case, the SDR method combined with the Gaussian randomization can recover a high-quality feasible solution of problem. Finally, we summarize the proposed  optimization scheme in Algorithm 2.

\subsection{Computational Complexity}

In this section, we will evaluate the computational complexity of the proposed algorithm without Gaussian randomization \cite{Kunyu}.
For the sub-problem (28), the iteration number is in the order of  $\sqrt {2LM} $,
then, there are $LM$ linear matrix inequality (LMI) constraints of size 1. Therefore, 
the complexity of problem (28) is in the order of $\sqrt {2LM} ( {LM(LM) + {{(LM)}^2}(LM) + {{(LM)}^3}} )$. Also, for the sub-problem (38), the iteration number is $\sqrt {LM + 2{N_{\rm{IRS}}} + 1} $ ,
there are $N_{\rm{IRS}}$ LMI constraints of size $N_{\rm{IRS}}$, and 
$(LM+N_{\rm{IRS}}+1)$ LMI constraints of size 1. As such, the complexity of solving problem (38) is in the order of
$\sqrt {2{N_{\rm{IRS}}} + LM + 1} ({N_{\rm{IRS}}^2 ({N_{\rm{IRS}}^3 + LM + {N_{\rm{IRS}}} + 1)}  + N_{\rm{IRS}}^4} $\\${(N_{\rm{IRS}}^2 + LM + {N_{\rm{IRS}}} + 1) + N_{\rm{IRS}}^6} )$. 

According to the above analysis, the total complexity is ${\cal O}[\sqrt {2{N_{\rm{IRS}}} + LM + 1} 2(N_{\rm{IRS}}^6 + 2N_{\rm{IRS}}^5 + N_{\rm{IRS}}^3 + (N_{\rm{IRS}}^4 + N_{\rm{IRS}}^2)(LM + 1)) + \sqrt {2LM} {(2{(LM)^3} + LM)^2})]$.

\begin{table}[!ht] 
	\small
	\renewcommand\arraystretch{1.4}
	\centering 
	\caption{}
	\begin{tabular}{l|l}\hline \hline 
		\bf{Parameters} & \bf{Value }\\ \hline
		{Number of the BS antennas} & {${{N_{\rm{TX}}=32}}$}  \\ \hline
		{Number of the IRS elements}  & {${N_{\rm{IRS}}}=20$}   \\ \hline
		{Number of the RF chains/clusters}       & {${{N_{\rm{RF}}}=4}$/${L=4}$}   \\ \hline
		{Number of the users/Eave}               & {$M=6$/$K=1$}          \\ \hline
		{Central frequency}                      &{$f = 340~{{\rm{GHz}}} $ } \\ \hline
		{The bits quantized of phase shifters}   &{$B=4$ }          	   \\ \hline
		{Transmit antenna gain}                  &{${G_t} = 1$}  \\ \hline
		{Receive antenna gain} &\hspace{-1.5ex} { ${G_r} = 4 + 20{\log _{10}} {\sqrt {N_{\rm{{TX}}}}}$} \\ \hline
		{Absorption cofficient } & {$\tau=0.0033/\rm{m}$ }         \\ \hline
		{Noise variance }&\hspace{-0.5ex}
		{${\sigma}^2=0.01{\rm{W}}$}    \\ \hline
		{Number of phase shifters of FC} & {${N_{\rm{PSF}}} = {N_{\rm{TX}}} {N_{\rm{RF}}}$}             \\ \hline
		{Number of phase shifters of SC} & {${N_{\rm{PSS}}} = {N_{\rm{TX}}}$ } \\ \hline\hline
	\end{tabular}
\end{table}

\section{Simulation Results}
In this section, we illustrate the performance of the proposed
IRS-aided THz MIMO-NOMA system. Due to the serious attenuation of THz propagation and absorption by water molecules, the scenario of short-distance communication is considered. We set the distances from BS to IRS and Eve are 15m and 5m, respectively. Moreover, the distance form IRS to all users and Eve is 5m, where users of per-cluster are located within a circle with 2m radius.
In the simulations, $M$ users are grouped into $N_{\rm{RF}}$ beams, and there are more than one user in each group \cite{Linglong1}. According to \cite{Gao1}, the SEE is defined as the ratio between the achievable SSR and the total power consumption 
\begin{equation}\label{44}
SEE = \frac{{\sum\limits_{l = 1}^L {\sum\limits_{m = 1}^M {R_{l,m}^{sec}} } }}{{\sum\limits_{l = 1}^L {\sum\limits_{m = 1}^M {{p_{l,m}} + } } {P_{\rm{RF}}}{N_{\rm{RF}}} + {P_{{\rm{PS}}}}{N_{\rm{PS}}} + {P_{\rm{B}}}}}.
\end{equation}
where $P_{\rm{RF}}$ is the power consumed by each RF chain, $P_{\rm{PS}}$ denotes
the power consumption of each phase shifter, and $P_{\rm{B}}$ is
the baseband power consumption. We set the
typical values $P_{\rm{RF}}= \rm{300~mW} $ , $P_{\rm{PS}}=\rm{40~mW} $ (4-bit phase
shifter), and baseband power $P_B= \rm{200~mW}$ \cite{Gao1}. $N_{\rm{PSF}}$ and  $N_{\rm{PSS}}$ are the number of phase shifters of FC and SC, respectively. Other simulation parameters are listed in TABLE I.

In addition, the following two benchmark schemes are compared with the proposed continuous phase shifts (CPS) scheme.

1. Discrete phase shifts (DPS): The initial IRS phase shifts are constructed by phase quantization, assuming that $\theta _{n}$ has $ \kappa $ discrete values, which are arranging evenly on the circle ${\bm{\theta}} = {e^{j{\theta _n}}}$\cite{Niu2}. Then, we have
\begin{equation}\label{45}
{{\bm{\theta}} _n} = \left\{ {{e^{j\frac{{2\pi \kappa}}{{{2^L}}}}}: \kappa = 0,1, \ldots ,{2^L} - 1} \right\}.
\end{equation}
where $L$ denotes the number of discrete binary levels.

2.~Random phase shifts (RPS): Only IRS is deployed, and the  phase shifts of reflection units are not optimized.

\begin{figure}[!htbp]
	\centering
	\includegraphics[width=9cm,height=8cm]{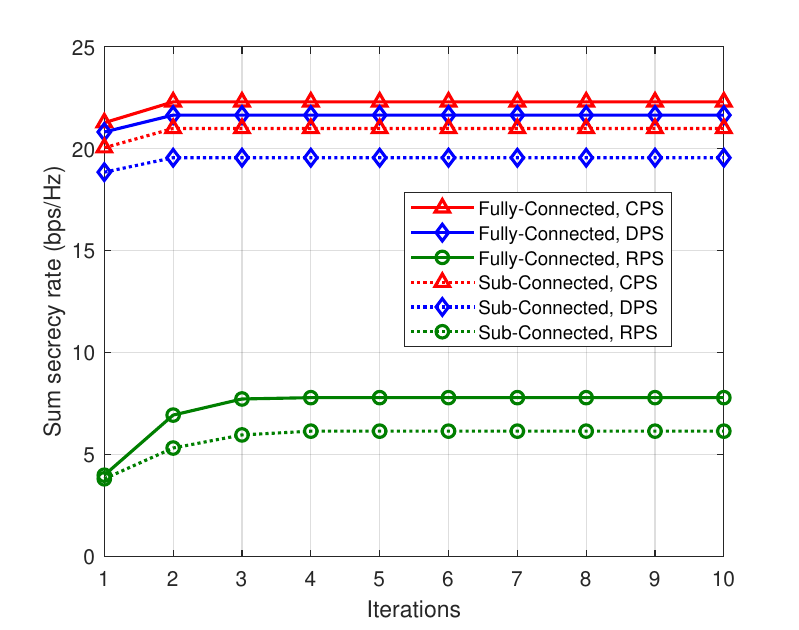}
	\caption{Sum secrecy rate versus the iteration numbers.}
	\label{fig:3}
\end{figure}

\begin{figure}[!htbp]
	\centering
	\includegraphics[width=9cm,height=8cm]{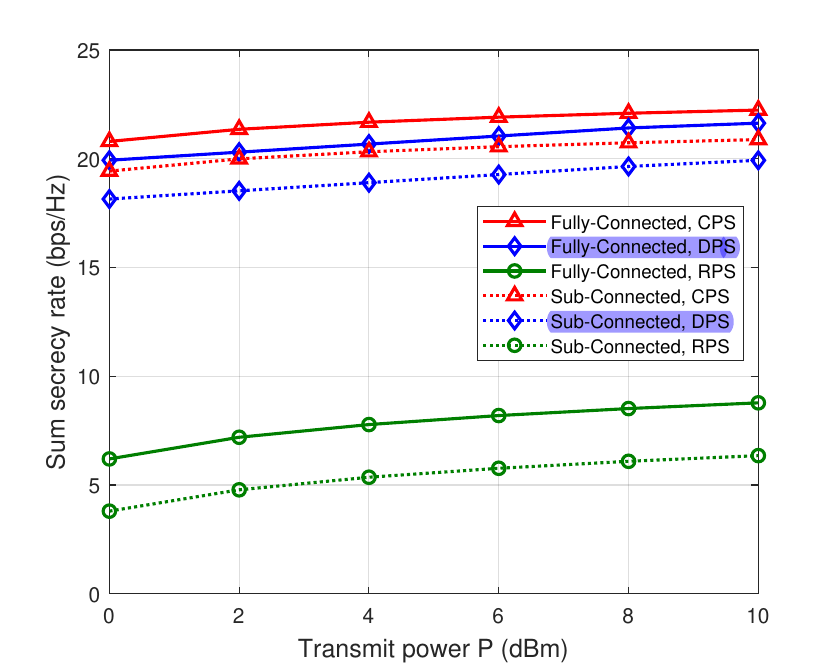}
	\caption{Sum secrecy rate versus the transmit power.}
	\label{fig:4}
\end{figure}
\begin{figure}[!htbp]
	\centering
	\includegraphics[width=9cm,height=8cm]{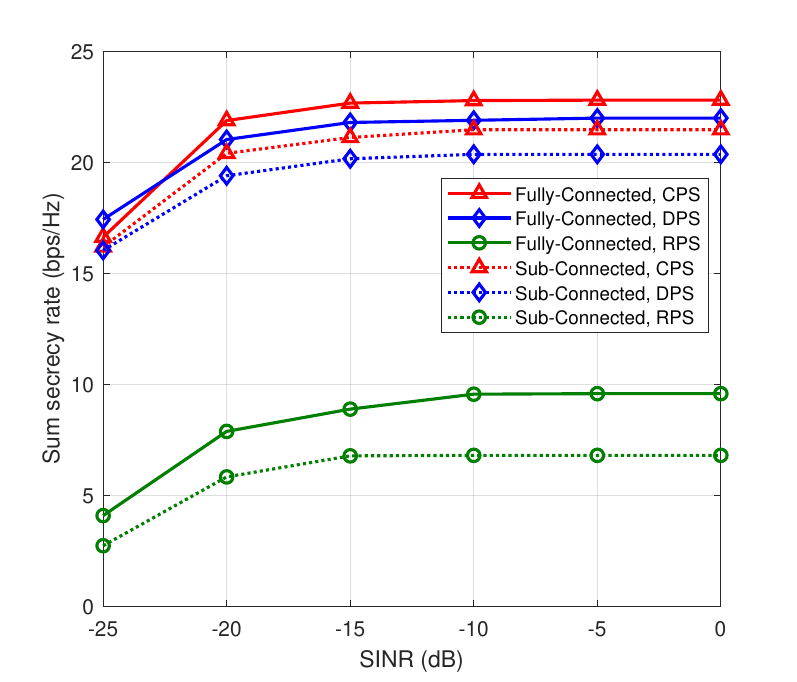}
	\caption{Sum secrecy rate versus the SINR of all users.}
	\label{fig:5}
\end{figure}

Fig. 3 illustrates the convergence performance of the proposed algorithm and benchmark schemes with the minimal SINR of the users, i.e., $R_{l,m}^{\min }=0~\rm{dB}$. It can be seen that all schemes can achieve quickly convergence, and the convergence speed of FC structure is the same as SC structure.
In addition, the iteration numbers of the RPS scheme is equal to inner loop iterations of the proposed method, which takes more than 5 iterations to reach convergence.
On the other hand, we can observe that the SSR of the FC structure is higher than that of the SC structure under the same conditions, which can be explained by the fact the full array gain can be utilized by each RF chain in the FC structure.


We compare the performance of two HP structure with respect to the BS transmit power in Fig. 4. As observed, the SSR of FC and SC structures increase monotonically with the increase of the transmit power. With the same HP structure, the achievable SSR after optimized IRS reflection phase shifts are significantly better than that of RPS schemes. This is due to the fact that the nature of IRS is similar to multi-antenna technology, which can bring new spatial degrees of freedom (DoF) and diversity gain to the system. Optimizing the phase shifts of IRS can converge the energy of multi-path signals and reflect it to the legitimate receivers, and suppress the eavesdropping, thus effectively improving the secrecy rate of the system.
What is more, in the practical application of IRS, discrete phase has more engineering significance and cost-effectiveness. However, there is a certain phase ambiguity at the IRS in the process of reflecting signals, which leads to the partial loss of performance. Therefore, the achievable SSR of DPS scheme is slightly lower than that of CPS.

Fig. 5 describes the relationship between the SSR and the minimal SINR requirement of users. As clearly shown by the results, the SSR of all schemes increase monotonously with the SINR threshold. It is worth noting that the growth trend of the SSR gradually becomes gentle with the increase of the SNIR requirement due to the limitation of the total transmit power budget.
The FC structure of each scheme can achieve better performance, this is due to fact that the all $N_{\rm{RS}}$ chains are used to serve for the users, which can bring full use of the multiplexing gain.
In addition, with the same SNIR condition, the passive beamforming gain brought by optimizing the phase shifts of IRS elements makes the scheme of CPS and DPS significantly better than that of RPS. Furthermore, the performance of DPS scheme is slightly worse than that of RPS scheme.
\begin{figure}[!htbp]
	\centering
	\includegraphics[width=9cm,height=8cm]{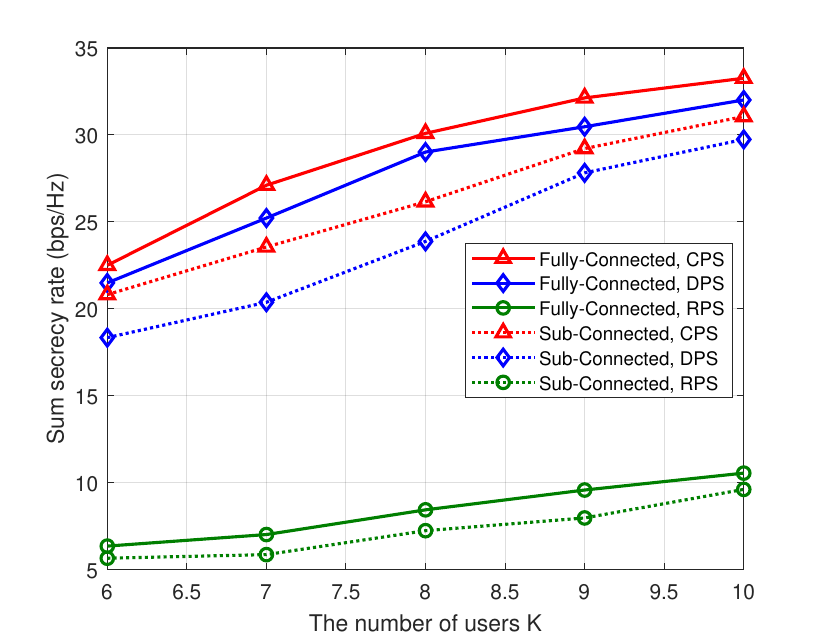}
	\caption{Sum secrecy rate versus the the number of users $K$.}
	\label{fig:6}
\end{figure}
\begin{figure}[!htbp]
	\centering
	\includegraphics[width=9cm,height=8cm]{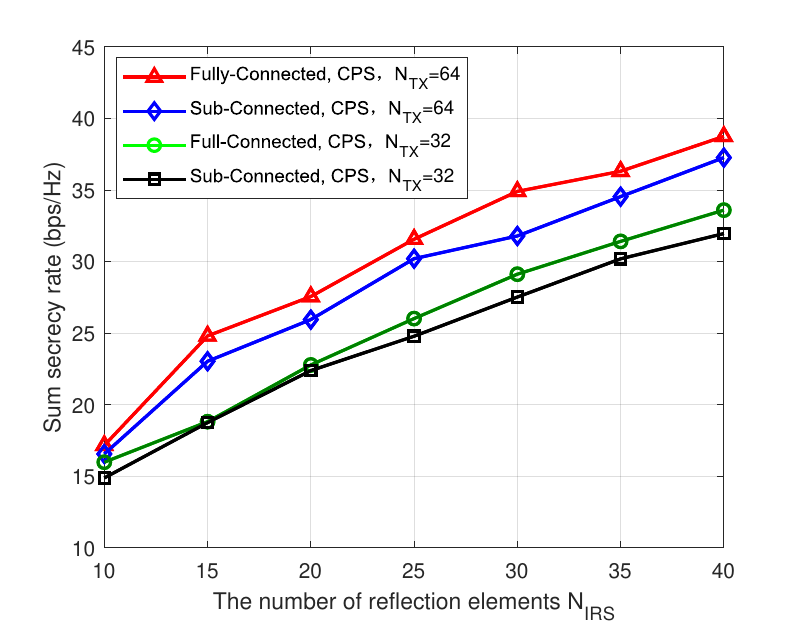}
	\caption{Sum secrecy rate versus the number of IRS reflection elements.}
	\label{fig:7}
\end{figure}

\begin{figure}[!htbp]
	\centering
	\includegraphics[width=9cm,height=7.8cm]{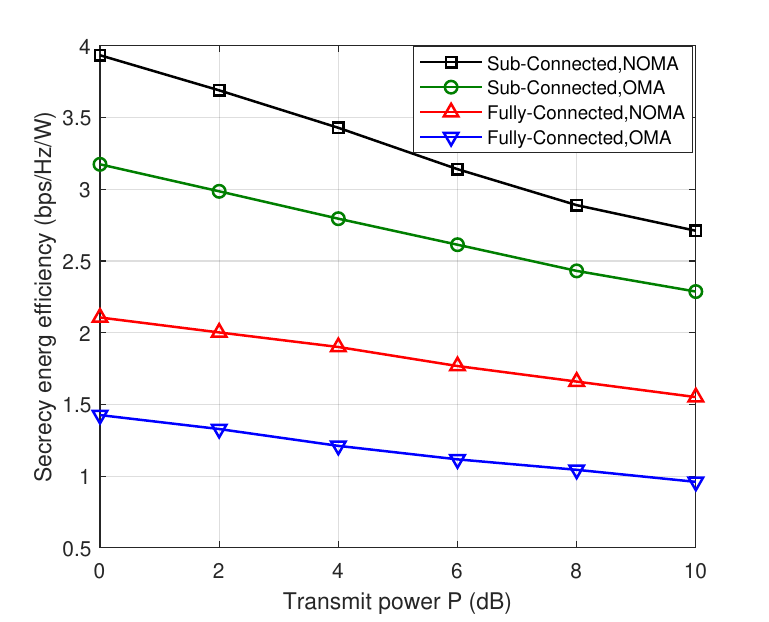}
	\caption{Secrecy energy efficiency versus the transmit power.}
	\label{fig:8}
\end{figure}

In Fig. 6, we plot the achievalbe SSR against the number of users $K$. Obviously, the results indicate that the achievable SSR of all schemes increases with the number of users. However, the growth trend of the RPS scheme is gentler than that of the CPS and DPS. The reason is that optimizing the IRS reflection coefficients can enhance the directional beamforming gains of reflection. In addition, in the proposed NOMA network, the reflected signal from the IRS serves all users at the same time, which can produce partial multiplexing gains.

In Fig. 7, we show the achievable SSR against the different numbers of IRS reflection elements  $N_{\rm{IRS}}$. 
It is intuitive that a larger number of IRS reflection elements will result in a higher SSR. This is due to the fact that the proposed scheme can take full advantages of the IRS to improve the channel conditions, i.e. the additional spatial DoF and passive beamforming gain provided by the IRS will gradually increase with the number of reflection elements. 
In addition, the results show that a higher beamforming gain can be achieved when there are more antennas at the BS. It is because that more antennas can further excavate the spatial dimension resources, so that multi-users can use the spatial freedom provided by MIMO to communicate with the BS on the same time frequency resources, thereby improving the secure spectrum efficiency without increasing the density of the BS.

We compare the SEE of NOMA with that of the conventional OMA scheme are provided in Fig. 8. According to the results, we can observe that SEE of all schemes is slowly degrading with the increases of transmit power. 
This can be explained by the fact that the power consumption for increasing the SSR of the system is larger than the benefits that it brings to the system, which causes the SEE of the system is reduced. Therefore, we should weigh the two performance indicators of SEE and secrecy rate as needed in practical. In addition, it can be verified that the proposed NOMA can obtain a higher SEE than OMA with the same HP structure. Meanwhile, the SC structure can achieve higher SEE than that of the FC structure. The reason is that the number of phase shifters used in SC structure is less than FC structure, which can save some power overhead.

\section{Conclusion}
In this paper, a secure THz MIMO-NOMA system via IRS was studied, where the users were grouped into multiple clusters based on channel correlation, and each cluster is served by a beam through NOMA. Moreover, a HP scheme with both FC and SC architectures were considered, where analog precoding was designed based on discrete phase and digital precoding was designed by the ZF technique.
We aimed to maximize the SSR by jointly optimizing power allocation and IRS phase shifts, while satisfying the constraints of transmit power, achievable rate requirement of users, and constant modulus of IRS. However, it is challenging to solve this problem directly due to the its non-convexity and coupled variables.
To this end, the original problem was decomposed into two sub-problems, which can be converted into the convex form by SDR and the first-order Taylor expansion. Then, with the obtained analog and digital precoding, the feasible solution can be acquired via the AO algorithm.
Finally, simulation results illustrated the proposed algorithm can converge quickly, and the passive beamforming of IRS can significantly improve security. In addition, FC can obtain the higher SSR than SC, but with a lower SEE. Thus, there was a compromise between SEE and the SSR of the proposed NOMA scheme.

The IRS integrated by passive components does not have the ability to process baseband signals, and cannot accurately obtain eavesdropping CSI. In addition, passive IRS can cause multiplicative fading effect, resulting in serious performance loss. The active IRS is equipped with a low-cost power amplifier, which is beneficial to obtain CSI and improve the reflection beamforming gain. Therefore, we will study the secure transmission of active IRS assisted THz systems in the future work.

\section*{Appendix A \\ The proof of theorem 1}
In the $t_1$-th iteration, we can obtain the following inequalities
\begin{equation}\label{46}
{\rm{lo}}{{\rm{g}}_2}\left( {B_1^{({t_1})}} \right) \le {\log _2}(\overline B _1^{({t_1})}) + \frac{{{\rm{Tr}}\left( {\overline {\bf{G}} _{l,m}^{}{{\bf{\Omega} }^{({t_1})}}\overline {\bf{G}} _{l,m}^H} \right)}}{{\overline B _1^{({t_1})}\ln 2}},
\end{equation}
\begin{equation}\label{47}
{\rm{lo}}{{\rm{g}}_2}\left( {C_1^{({t_1})}} \right) \le {\log _2}(\overline C _1^{({t_1})}) + \frac{{{\rm{Tr}}\left( {\overline {\bf{G}} _{l,m}^E{\bf{\Psi }}_{}^{({t_1})}\overline {\bf{G}} {{_{l,m}^E}^H}} \right)}}{{\overline C _1^{({t_1})}\ln 2}},
\end{equation}
where 
\begin{equation}\label{48}
{{\bf{\Omega} }^{({t_1})}}\hspace{-0.5ex} =\hspace{-0.5ex} {\left| {{{\bf{v}}_l}} \right|^2}\sum\limits_{j = 1}^{m - 1} {\left( {{p_{i,j}^{({t_1})}}\hspace{-0.5ex} -\hspace{-0.5ex}  p _{i,j}^{({t_1-1})}} \right)}\hspace{-0.5ex}  + \hspace{-0.5ex}\sum\limits_{i \ne l}^L {{{\left| {{{\bf{v}}_i}} \right|}^2}\sum\limits_{j = 1}^M {\left( {{p_{i,j}^{({t_1})}} \hspace{-0.5ex}- \hspace{-0.5ex} p _{i,j}^{({t_1-1})}} \right)} }. 
\end{equation}
and ${\bf{\Psi }} \hspace{-0.7ex} =\hspace{-0.7ex} \sum\limits_{i = 1}^L \hspace{-0.7ex}{{{\left| {{{\bf{v}}_i}} \right|}^2}\sum\limits_{j = 1}^M {\left( {{p_{i,j}^{({t_1})}} \hspace{-0.5ex}- \hspace{-0.5ex}  p _{l.j}^{({t_1-1})}} \right)} }.$
Due to the linear approximation method is utilized in (45)-(47), the update rule in algorithm 1 can ensure the solution $ p_{l,m}^{({t_1})}$ is a optimal solution. As such, the inequality $f(t_1) \ge f(t_1-1)$ holds. Algorithm 1 can generate a non-decreasing sequence of the objective value. The objective function of the problem (28) is bounded due to the total power budget constraint and the minimum achievable rate requirement of  users. Therefore, algorithm 1 can guarantee convergence to a local optimal solution, i.e., $\left\| {{\rm{\Delta }}l} \right\| = 0$ and $f(t_1) = f(t_1-1)$.

\bibliographystyle{ieeetr}
\setlength{\baselineskip}{12pt}
\bibliography{reference}

\ifCLASSOPTIONcaptionsoff
  \newpage
\fi

\end{document}